# Mode-shape perturbation induced by analyte adsorption in nanomechanical sensors

Mert Yüksel[1] and M. Selim Hanay[1,2,*]

[1]*Department of Mechanical Engineering, Bilkent University, Ankara 06800, Turkey*
[2]*National Nanotechnology Research Center (UNAM), Bilkent University, Ankara 06800, Turkey*

E-mail: selimhanay@bilkent.edu.tr

Nanomechanical resonators offer important benefits for the sensing of physical stimuli such as the mass of an added molecule. To map out the local shape properties of the physical stimuli, such as the distribution of the mass density of a molecule, sensory information should be collected through multiple modes of a mechanical sensor. By utilizing the specific mode shapes, the spatial distribution of a physical stimulus can be reverse calculated. However, the mode shapes of a sensor may deviate from their ideal forms once analytes start to accumulate on the sensor. As a result, algorithms based on the ideal form of the mode shapes no longer work accurately. Here, we developed a theoretical framework to calculate the change in the mode shapes of a nanomechanical beam after analyte adsorption. We verified the theoretical model by performing finite element simulations and comparing the change in the mode shapes obtained from each approach. Monte Carlo simulations were performed to relate the maximum deviation in the mode shapes to the surface roughness of the sensor after analyte accumulation. By predicting the change in the mode shapes and using the corrected forms, the accuracy of the nanomechanical sensing can be improved significantly.

# I. INTRODUCTION

Nanoelectromechanical system (NEMS) resonators can be used as exquisite sensors of physical stimuli. As the size of a mechanical sensor shrinks, its responsivity increases. Combining this advantage with low-noise readout schemes has enabled extreme sensitivity levels to be achieved, such as the detection of electronic spins [1], single-atoms [2], single-proteins [3,4], nanoparticles [5,6], single-cells in liquid [7-9] and volatile chemicals at low concentration [10,11]. A novel sensing modality of NEMS technology for single-molecule analysis has been discovered recently. It was shown that the spatial properties of analyte particles, such as their size and skewness, can be extracted by using multiple mechanical modes of a sensor [12]. Such multimodal measurements provide both size and shape features, as well as the mass of the analyte. Furthermore, by merging different spatial features, an image of the analyte can be reconstructed. The new technique, inertial imaging [12], transforms the capabilities of nanomechanical sensors to a new level: the combined data of molecular mass, size and shape of the analyte can provide previously unattainable insights in biomolecular analytics. The technique can be used for any resonator system with well-defined mode shapes: for instance, the electromagnetic analog of inertial imaging was also demonstrated in a microfluidic environment using microwave sensors [13].

Inertial imaging and similar multimode techniques [14-23] work on the principle that the mode shapes of the sensor are well known. However, the mode shapes in a real device deviate from the ideal mode shapes since there are many factors (surface effects, clamping conditions, fabrication errors) that can affect the physical composition of sensors. To apply inertial imaging algorithms, such deviations from ideality should be taken into account.

Here, we take the first step in doing so by quantifying the change in the mode shape of a NEMS resonator after analyte adsorption. We model an analyte by considering its inertial effect on the resonator: the contribution to the elastic energy is ignored because Young's moduli of biological material are usually orders of magnitude smaller than those of crystalline material used for NEMS. However, the approach provided here can be generalized for such cases in which the contribution of the analytes to the stiffness is important. First, we used first-order perturbation theory to obtain the change in resonance frequency and mode shapes. Then, we performed simulations to verify that these theoretical predictions were correct. We demonstrated the utility of our approach by placing two particles on the NEMS sensor sequentially: the parameters of the second particle (mass and position) were obtained much more accurately once the effect of the first particle on mode shape was taken into account. We then used this theory to study the effect of surface roughness on mode shapes.

## II. THEORY

Consider a doubly-clamped beam extending along the $x$ direction. It has a Young's modulus $E$, cross sectional area $A$, mass density $\rho$ and moment of inertia $I$. The equation of motion for the $n^{th}$ flexural mode of this structure is [24]:

$$EI\frac{\partial^4}{\partial x^4}\phi_n^{(0)}(x) = \rho A\, \Omega_n^{(0)}\, \phi_n^{(0)}(x) \qquad (1)$$

Here, $\phi_n^{(0)}$ is the unperturbed mode shape and $\Omega_n^{(0)} = \omega_{n,0}^2$, in which $\omega_{n,0}$ is the unperturbed resonance frequency of the $n^{th}$ mode.

In a sensing application, an analyte gets adsorbed by this doubly clamped beam and changes the composition of Eq. (1). To account for the finite size and shape of an analyte, we consider its density profile along the beam axis $x$. The 1D density profile of this analyte is shown by $\epsilon \times \mu(x)$, in which $\epsilon \ll 1$ is introduced as the smallness parameter for this perturbation. The equation of motion after the perturbation of analyte becomes:

$$EI\frac{\partial^4}{\partial x^4}\phi_n(x) = \rho A\, \Omega_n\, \phi_n(x) + \epsilon\, \mu(x) A \Omega_n\, \phi_n(x) \qquad (2)$$

in which $\phi_n(x)$ and $\Omega_n$ are the updated values for the mode shape and the square of the resonance frequency, respectively. These parameters can be expressed as a perturbation series [25]:

$$\phi_n(x) = \phi_n^{(0)}(x) + \epsilon\, \phi_n^{(1)}(x) + \mathcal{O}(\epsilon^2) \qquad (3)$$

$$\Omega_n = \Omega_n^{(0)} + \epsilon\, \Omega_n^{(1)} + \mathcal{O}(\epsilon^2) \qquad (4)$$

in which $\epsilon\, \phi_n^{(1)}$ and $\epsilon\, \Omega_n^{(1)}$ are the first order corrections to the mode shape and the square of the resonance frequency, respectively. Noting that $\omega_n = \omega_n^{(0)} + \epsilon\, \omega_n^{(1)} + \mathcal{O}(\epsilon^2)$, the relationship between the first-order shift in resonance frequency, $\omega_n^{(1)}$, and the $\Omega_n^{(1)}$ can be determined:

$$\Omega_n^{(1)} = 2\, \omega_n^{(0)} \omega_n^{(1)} \qquad (5)$$

To proceed further, we substituted these expressions into the equation of motion, collected the terms at the same order in $\epsilon$, performed integration by parts, and invoked the boundary conditions to cancel out several terms, as explained in the Annex. In this way, we obtained the first-order correction to the frequency:

$$\frac{\omega_n^{(1)}}{\omega_n^{(0)}} = -\frac{1}{2\rho}\int dx\, \mu(x)\left(\phi_n^{(0)}(x)\right)^2 \qquad (6)$$

which is exactly the equation obtained by other methods in the literature [12]. We then turned our attention to the first-order correction to the mode shape, which had not been calculated before. We wrote the correction function as a superposition of the original mode shapes:

$$\phi_n^{(1)} = \sum_{m \neq n}^{\infty} c_m^n \phi_m^{(0)} \qquad (7)$$

Here, the right-hand side does not include the term $m = n$, because any such term (i.e., $c_n^n \phi_n^{(0)}$) would not constitute a deviation from the original mode shape ($\phi_n^{(0)}$) and can always be normalized out. Substituting in the formula shown in Eq. (7) into Eq. (2) and multiplying both sides with $\phi_m^{(0)}$ and integrating, we obtained the sought-after weights, $c_m^n$:

$$c_m^n = \frac{1}{\rho} \frac{\Omega_n^{(0)}}{\Omega_m^{(0)} - \Omega_n^{(0)}} T_{nm} \qquad (8)$$

in which $T_{nm}$ quantifies the triple-overlap between the two modes and the analyte:

$$T_{nm} \equiv \int dx\, \mu(x)\, \phi_n(x) \phi_m(x) \qquad (9)$$

Therefore, the first order correction to the mode shape is:

$$\phi_n^{(1)} = \frac{1}{\rho} \sum_{m \neq n} \frac{\Omega_n^{(0)}}{\Omega_m^{(0)} - \Omega_n^{(0)}} T_{nm}\, \phi_m^{(0)} \qquad (10)$$

### III. SIMULATIONS

To verify that the correction to the mode shapes calculated in Eq. (10) is accurate, we performed finite element simulations. We used COMSOL Multiphysics to simulate the dynamics of analyte adsorption on a doubly-clamped beam. Eigenfrequency analysis is performed with and without analyte adsorption to observe perturbed and unperturbed out-of-plane mode shapes and the corresponding resonance frequencies. Care was taken to normalize the mode shapes properly so that the orthonormality condition was satisfied for each mode. To model the analyte, a uniform density profile, $\mu(x)$, was defined between specified $x$-positions on the beam model, as shown in Fig. 1. Simulations were repeated with finer meshes until a convergence limit well beyond the effect of the analyte was reached. Exactly the same mesh specifications were defined for the beam, in both cases with and without the adsorbed mass, to get an accurate comparison. For the particular simulation that is shown in Fig. 1, the density profile, $\mu(x)$, was chosen to be less than the beam's material density to be able to observe small perturbations. It is centered on $1/4^{th}$ of the beam, and extends along the 5% of the beam's length. The first order correction to the mode shape was calculated for the first ($\phi_1^{(1)}$) and second ($\phi_2^{(1)}$) out-of-plane mode shapes from the simulations by simply taking the difference between the perturbed and unperturbed mode shapes. To compare this deviation with the theoretical prediction, Eq. (10) was evaluated up to the first 10 modes. A comparison between the numerical and theoretical results is shown in Fig. 1, which indicated that our analytical prediction matches well with the results from finite element analysis (FEA). Specifically,

Fig. 1(b) shows the first-order deviation of the first mode, which resembles the ideal shape of the second mode. In this particular case, the largest weight in Eq. (7) was $c_2^1$, which means that the correction to the mode shape of the first mode should resemble the ideal mode shape of the second mode. A similar situation holds true for the correction to the second-mode, as shown in Fig. 1(d), in which the largest contribution comes from the ideal shape of mode 1 followed by mode 3.

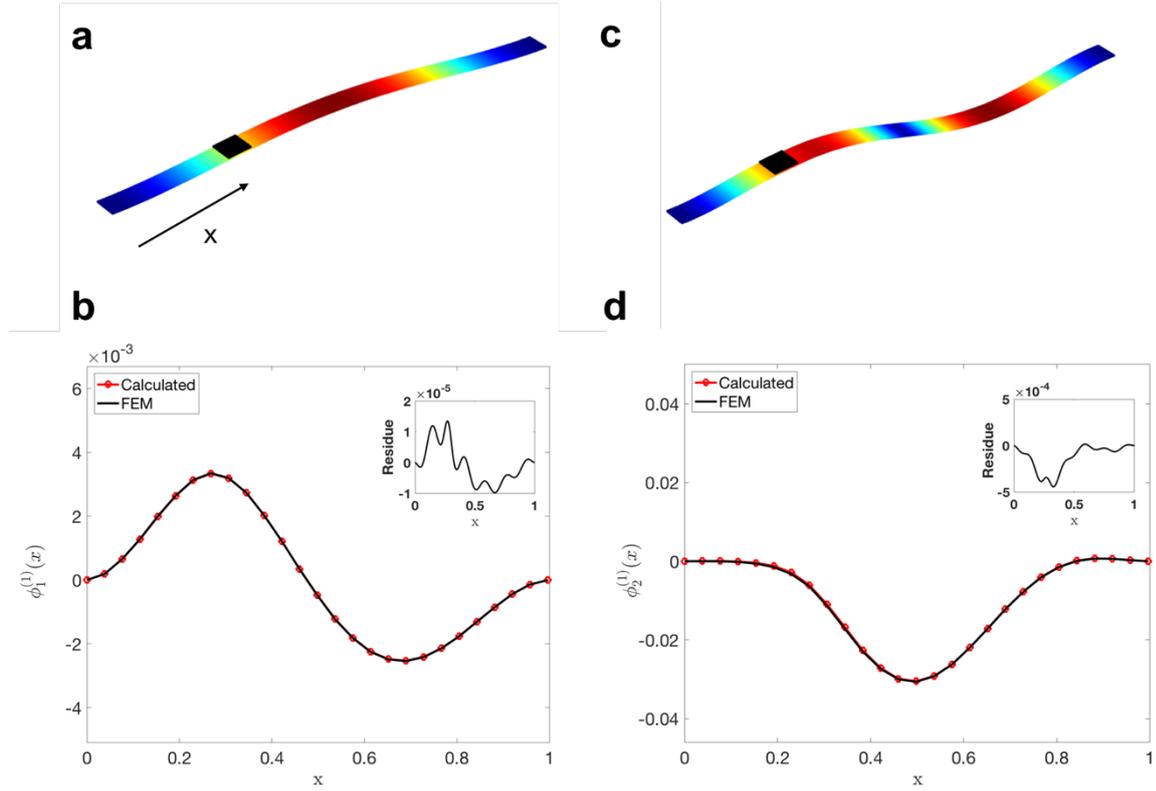

**FIG. 1.** *Performance of calculating the corrected mode shapes. (a) The mode shape of the first resonance of the doubly clamped beam used in finite element simulations with the adsorbed particle (black) located in position x. (b) The first-order correction to the mode shape for the **first mode** is calculated (red) and compared to the result obtained by FEM simulations (black). (c) The mode shape of the second resonance of the beam with the adsorbed particle. (d) The first-order correction to the mode shape for the **second mode** is calculated (red) and compared to the result obtained by FEM simulations (black).*

Once the deviation from the ideal mode shapes are calculated, the perturbed mode shapes can be reconstructed by:

$$\phi_n(x) = c_n^n \phi_n^{(0)} + \epsilon \sum_{m \neq n} c_m^n \, \phi_m^{(0)} \qquad (11)$$

The error between the estimated and FEA-perturbed mode shape is calculated in the following way:

$$\%Error = \frac{\|\phi_{n_{fea}} - \phi_{n_{estimated}}\|}{\|\phi_{n_{fea}}\|} \times 100 \tag{12}$$

The error is calculated for the different number of modes used in Eq. (12) [Fig. 2(a)]. For the first three modes, it is observed that the error decays with increasing number of modes used, and it saturates after 10 terms. To test the reliability of the estimation, the error was also calculated as a variation of the position of the adsorbed density profile. Figure 2(b) shows that the analytical technique works for a range of positions of the adsorbed particle.

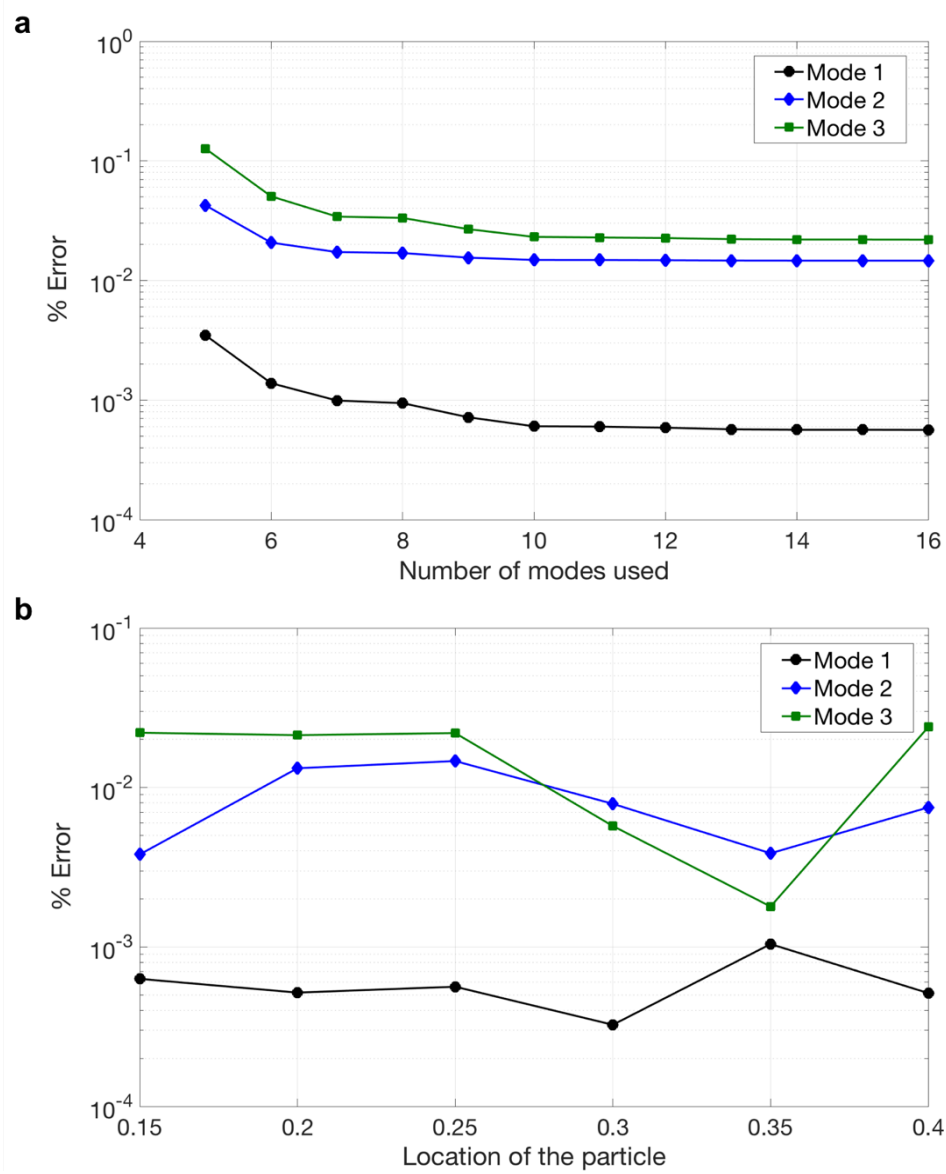

**FIG. 2.** *(a) Error vs. the number of modes used. As more modes are used in the model, much better agreement between the theoretical and FEM simulations is obtained. For the first three modes, the errors drop quickly, then saturate to a residual value. (b) Error vs. position of the adsorbed particle. The performance of the prediction is evaluated for a particle adsorbed at different locations on the beam. The error does not change significantly if the particle is located at different positions.*

So far, we have only considered the case in which a single particle has been adsorbed. Now, we locate a second particle identical to the first one in a way that they do not overlap. We studied how the perturbation of the first particle on the mode shapes changes the performance for detecting the properties of the second particle. We kept the position of the first particle ($x_1$) constant, and swept the position of the second particle ($x_2$) along the beam [Fig. 3(a)]. Then, we calculated $x_2$ by using both the normal (unperturbed) mode shapes and the corrected (perturbed) mode shapes after the first particle. The same procedure was applied for different $x_1$ locations, and the mean error for the estimated $x_2$ were calculated, as shown in Fig. 3(b). When we compared the calculated values with the actual center-of-mass position of the second particle, we observed that using corrected mode shapes for detection provided better accuracy for the center-of-mass estimation. The error using the original mode shapes may seem to be small already. However, the errors propagate when the higher order moments of a distribution are calculated; therefore, it is crucial to decrease the error in determining the position as much as possible. The procedure can be repeated by updating the corrected mode shape whenever a new particle is adsorbed, which enables more precise particle position tracking.

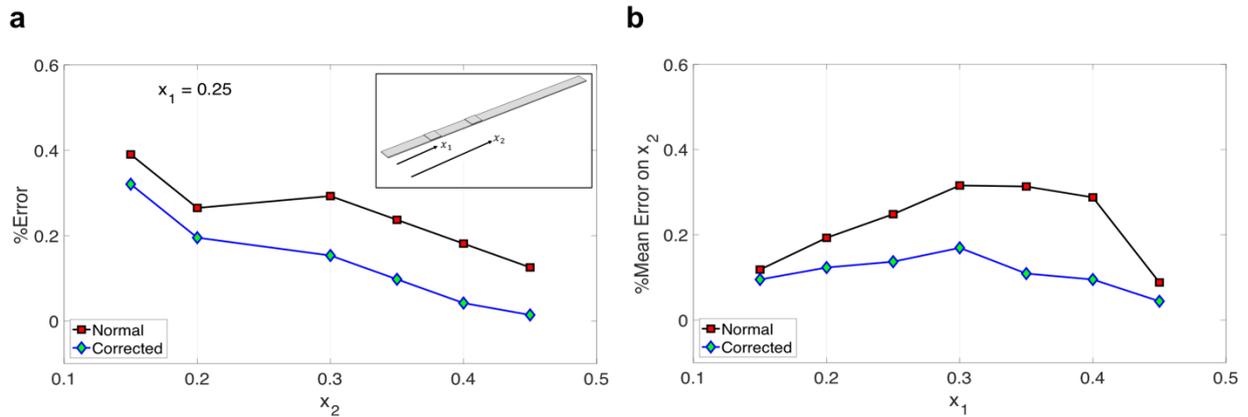

**FIG. 3.** *Position prediction accuracy. (a) The center-of-mass position of the first particle ($x_1$) is fixed at $x_1 = 0.25$ and the center-of-mass position of the second particle ($x_2$) is moved along the beam. For each $x_2$ values, the position prediction algorithm is used both with the normal mode shapes (black line), and the corrected mode shapes (blue line), which are the perturbed mode shapes after the adsorption of the first particle. The error between the calculated and actual positions is plotted. (b) The mean error of $x_2$ is calculated from (a) for different positions of the first particle ($x_1$) to test the reliability of our method.*

We can use this technique to work on more general problems, such as the effect of surface roughness on mode shapes. The effect of the surface roughness on the mode shapes of a nanomechanical beam is generally disregarded in NEMS applications. However, surface roughness can have a negative effect on the sensitivity of the application. Here, we applied our approach to determine the variation in the mode shapes of the nanomechanical beam due to surface roughness. For simplicity, we considered a case in which surface roughness is caused by soft adsorbates so that the elasticity contribution of the adsorbates can be ignored. In our simulations, we implemented roughness profile as particles attached to the surface of the beam. We created random surface roughness profiles, $z(x)$, by controlling the number of adsorbed particles and their height and length along the surface [Fig. 4(a)]. For each randomly generated surface profile, the roughness rms value was calculated in the following way:

$$Roughness\ rms = \sqrt{\frac{1}{L}\int_0^L z(x)^2 dx} \qquad (13)$$

in which $L$ is the length of the beam, which is normalized to 1 unit. Then the deviation from the normal mode shape was calculated:

$$\%Deviation, Mode\ Shape\ 1 = \frac{\left\|\phi_1^{(0)} - \phi_1\right\|}{\left\|\phi_1^{(0)}\right\|} x100 \qquad (14)$$

in which $\phi_1$ represents the calculated mode shape with the surface roughness. Figure 4(b) shows the deviation in the first mode shape depending on the surface roughness. Although for some cases, the mode shape does not change even for a large surface roughness, there is a clear trend that, on average, a larger surface roughness implies a larger deviation in the mode shape.

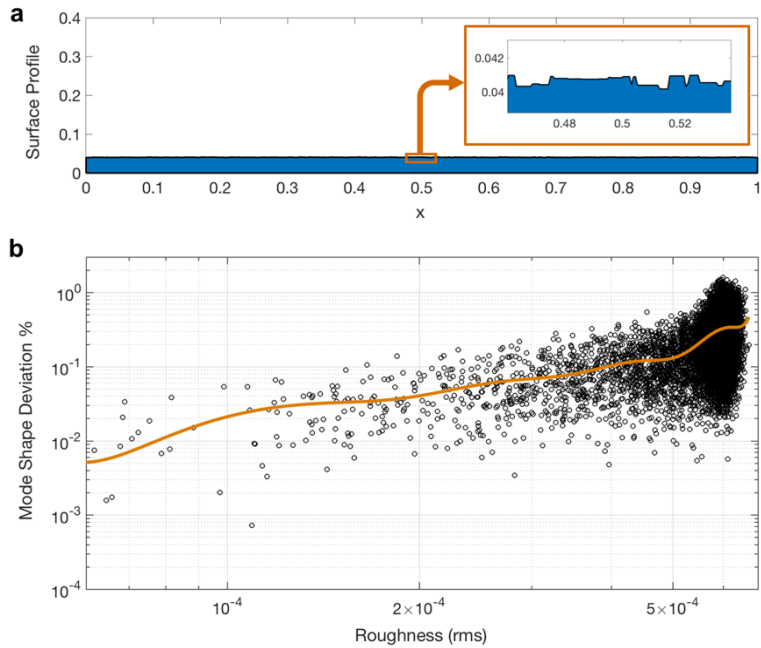

***FIG. 4.*** *(a) Surface roughness profile of the beam in one of the Monte Carlo simulations. The inset show a close-up view of the surface after adsorption of analytes. (b) Surface roughness vs. the deviation in the first mode shape. The orange line shows the trend line obtained by fitting a 12*[th] *degree polynomial curve to the data.*

## IV. SUMMARY / CONCLUSION

In conclusion, we derived and verified a mathematical procedure to estimate the perturbed mode shapes of a nanomechanical beam after the adsorption of a species. We used our mathematical procedure for a doubly clamped nanomechanical beam resonator and showed that it agrees with the results from finite element analysis. Then, we showed that using perturbed mode shapes increases the sensitivity for mass position sensing. NEMS sensing applications highly depend on the change in the resonance frequencies and the mode shapes. Although it is straightforward to detect the change in the resonance frequencies, there is no easy way to sense the change in the mode shapes. Therefore, this change is neglected in many NEMS applications. However, new techniques such as inertial imaging depends on the mode shapes. Here, we show that perturbed mode shapes of a nanomechanical beam can be easily calculated. When these changes are taken into account, the sensitivity of the NEMS application is improved. The approach provided here can be used to record a *history* of particles adsorbed on a structure and update the mode shapes accordingly so that spatial sensing can be performed accurately.

## ACKNOWLEDGEMENTS

MS Hanay acknowledges financial support from the European Commission, FP7 Marie Curie Program in the form of a Career Integration Grant number 631728.

APPENDIX

Continuing from Eq. (5) in main text, the expression after plugging in the perturbation expansion reads:

$$EI \frac{\partial^4}{\partial x^4} \phi_n^{(1)}(x) - \mu(x) A \Omega_n^{(0)} \phi_n^{(0)}(x) = \rho A \, \Omega_n^{(0)} \phi_n^{(1)}(x) + \rho A \, \Omega_n^{(1)} \phi_n^{(0)}(x) \tag{A1}$$

To proceed further, we multiply both sides of the with $\phi_n^{(0)}(x)$ and integrate:

$$\int dx \, \phi_n^{(0)}(x) \, EI \frac{\partial^4}{\partial x^4} \phi_n^{(1)}(x) - \int dx \, \phi_n^{(0)}(x) \, \mu(x) A \Omega_n^{(0)} \phi_n^{(0)}(x)$$
$$= \int dx \, \phi_n^{(0)}(x) \, \rho A \, \Omega_n^{(0)} \phi_n^{(1)}(x) \tag{A2}$$
$$+ \int dx \, \phi_n^{(0)}(x) \, \rho A \, \Omega_n^{(1)} \phi_n^{(0)}(x)$$

Applying integration by parts to the first term on the left hand side and using the boundary conditions on each end ($x=0$ and $x=1$) of the doubly-clamped beam: $\phi_n^{(0)}(0,1) = \phi_n^{(1)}(0,1) = \frac{\partial \phi_n^{(0)}}{\partial x}\Big|_{x=0,1} = \frac{\partial \phi_n^{(1)}}{\partial x}\Big|_{x=0,1} = 0$, we obtain:

$$\int dx \, \phi_n^{(0)}(x) \, EI \frac{\partial^4}{\partial x^4} \phi_n^{(1)}(x) = \int dx \, \phi_n^{(1)}(x) \, EI \frac{\partial^4}{\partial x^4} \phi_n^{(0)}(x)$$
$$= \int dx \, \phi_n^{(1)}(x) \, \rho A \Omega_n^{(0)} \phi_n^{(0)}(x) \tag{A3}$$

which cancels out the first term on the right hand side of Eq. (7).

The second term on the right hands side simply integrates to $\rho A \, \Omega_n^{(1)}$. Therefore, Eq. (7) now reads:

$$-\Omega_n^{(0)} \int dx \, \phi_n^{(0)}(x) \, \mu(x) \, \phi_n^{(0)}(x) = \rho \, \Omega_n^{(1)} \tag{A4}$$

Converting back to frequency shifts, we obtain the first order correction to resonance frequency:

$$\frac{\omega_n^{(1)}}{\omega_n^{(0)}} = -\frac{1}{2\rho} \int dx \mu(x) \left(\phi_n^{(0)}(x)\right)^2 \tag{A5}$$